\documentclass[12pt]{iopart}
\usepackage{iopams}
\usepackage{graphicx}
\begin{document}

\title[]{Magnetization of multicomponent ferrofluids}
\author{I. Szalai$^1$ and S. Dietrich$^{2,3}$}

\address{$^1$Institute of Physics and Mechatronics, University of Pannonia, H-8201 Veszpr\'em, PO Box 158, Hungary\\
$^2$Max-Planck-Institut f\"ur Intelligente Systeme, Heisenbergstr. 3, D-70569 Stuttgart, Germany\\
$^3$Institut f\"ur Theoretische und Angewandte Physik, Universit\"at Stuttgart, Pfaffenwaldring 57, 
D-70569 Stuttgart, Germany}
\ead{szalai@almos.vein.hu, dietrich@mf.mpg.de}
\begin{abstract}
The solution of the mean spherical approximation (MSA) integral equation for isotropic multicomponent dipolar hard sphere fluids without external fields is used 
to construct a density functional theory (DFT), which includes external fields, in order to obtain an analytical expression for the external field dependence of the 
magnetization of ferrofluidic mixtures. 
This DFT is based on a second-order Taylor series 
expansion of the free energy density functional of the anisotropic 
system around the corresponding isotropic MSA reference system. 
The ensuing results for the magnetic properties are in quantitative agreement with our canonical ensemble Monte Carlo simulation data presented here.
\end{abstract}

\section{Introduction}
Ferrofluids are colloidal suspensions of single domain ferromagnetic grains dispersed in a solvent. 
The stabilization of such suspensions is usually obtained by coating the magnetic particles with polymer or 
surfactant layers or by using electric double layer formation. 
Since each particle of a ferrofluid possesses a permanent magnetic dipole moment, upon 
integrating out the degress of freedom of the solvent, which gives rise to effective pair potentials, dispersions of ferrocolloids can be considered as paradigmatic realizations of dipolar liquids \cite{hl}. 
The effective interactions of such magnetic particles are often modeled by dipolar 
hard-sphere (DHS) \cite{ik,bi}, dipolar Yukawa \cite{sd}, or Stockmayer \cite{rd} interaction potentials. 
The most frequently applied methods to describe ferrofluids encompass mean field theories \cite{de,sdo}, 
thermodynamical perturbation theory \cite{ik}, integral equation theories \cite{mbc,ml,kf}, 
various DFTs \cite{tt,fd,gd1,gd2,sd2,sd}, as well as Monte Carlo \cite{ks1,tk} and 
molecular dynamics \cite{hw,iv,jk} simulations.
 
Within the framework of DFT and the mean spherical approximation (MSA), previously we have 
proposed an analytical equation \cite{sd} for the magnetic field dependence of the magnetization of 
{\it one}-component ferrofluids, which turned out to be reliable as compared with corresponding 
Monte Carlo (MC) simulation data. For this kind of system the effect of an external magnetic 
field has been taken 
into account by a DFT method, which approximates the free energy functional of 
the anisotropic system with an external field by 
a second-order Taylor series expansion around the corresponding isotropic reference 
system without an external field. The expansion coefficients are the 
direct correlation functions which for the studied isotropic dipolar hard-sphere (DHS) and 
dipolar Yukawa reference systems can be obtained analytically from Refs. \cite{we,he}.
 
However, in practice the magnetic colloidal suspensions are often multicomponent. 
In order to describe the magnetization of  ferrofluidic {\it mixtures} we extend 
our one-component theory to multicomponent systems. This extension is based on 
the multicomponent MSA solution obtained by Adelman and Deutch \cite{ad}. 
They showed that the properties of equally sized 
hard spheres with different dipole moments can be expressed in terms of those of an 
effective single component system. 
Because MSA is a linear response theory Adelman and Deutch could predict only the initial slope 
(or zero-field susceptibility) of the magnetization curve. Using their 
MSA solutions as those of a reference system, in the following we present DFT calculations of 
the {\it full} magnetization curves of equally sized, dipolar hard sphere mixtures. 
These results are compared with MC simulation data.      

\section{Microscopic model and MSA solution}
We consider dipolar hard-sphere (DHS) fluid mixtures which consist of $C$ components. 
The constitutive particles have 
the same diameter $\sigma$ but the strength $m_a$ of the embedded point dipole 
can be different for the components $a=1,...,C$.
In the following the indices $a$ and $b$ refer to the components while the 
indices $i$ and $j$ refer to individual particles.
The system is characterized by the following pair potential:
\begin{equation}
w_{ij}^{DHS}({\bf{r}}_{12},\omega_1,\omega_2)=w^{HS}_{ij}(r_{12})+w^{DD}_{ij}({\bf{r}}_{12},\omega_1,\omega_2),
\end{equation}
where $w^{HS}_{ij}$ and $w^{DD}_{ij}$ are the hard-sphere and the dipole-dipole interaction pair potential, respectively.
The hard-sphere pair potential given by
\begin{equation}
w^{HS}_{ij}(r_{12})=\left\{
        \begin{array}{lll}
     \infty &, & r_{12} < \sigma\\
     0      &, & r_{12} \geq \sigma.
        \end{array} 
        \right. \
\label{hs}
\end{equation}
The dipole-dipole pair potential is
\begin{equation}
w^{DD}_{ij}({\bf{r}}_{12},\omega_1,\omega_2)=-\frac{m_im_j}{r_{12}^3}D(\omega_{12},\omega_1,\omega_2),
\end{equation}
with the rotationally invariant function
\begin{equation}
D(\omega_{12},\omega_1,\omega_2)=
3[\widehat{\mathbf{m}}_1(\omega_1)\cdot\widehat{\mathbf{r}}_{12}]
[\widehat{\mathbf{m}}_2(\omega_2)\cdot\widehat{\mathbf{r}}_{12}]
-[\widehat{\mathbf{m}}_1(\omega_1)\cdot\widehat{\mathbf{m}}_2(\omega_2)],
\label{D}
\end{equation}
where particle 1 (2) of type $i$ ($j$) is located at ${\mathbf{r}}_1$ (${\mathbf{r}}_2$) and carries a 
dipole moment of strength $m_i$ ($m_j$) with an orientation given by the unit vector 
$\widehat{\mathbf{m}}_1(\omega_1)$ ($\widehat{\mathbf{m}}_2(\omega_2)$) with polar angles 
$\omega_1=(\theta_1,\phi_1)$ ($\omega_2=(\theta_2,\phi_2)$); 
${\mathbf{r}}_{12}={\mathbf{r}}_1-{\mathbf{r}}_2$ is the difference vector between 
the center of particle 1 and the center of particle 2 with $r_{12}=|{\mathbf{r}}_{12}|$.

Within the framework of MSA Adelman and Deutch \cite{ad} presented 
an analytical solution for the aforementioned $C$-component {\it isotropic} dipolar fluid mixture 
in the absence of external fields.
The importance of their contribution is that it provides simple analytic expressions for 
correlation functions, the dielectric constant (in our case the zero-field magnetic susceptibility), 
and thermodynamic functions. In our envisaged DFT calculations for dipolar mixtures {\it with} 
external fields we consider the isotropic DHS fluid mixture without external field as a 
reference system which is described by the following MSA second-order direct 
correlation function:  
\begin{eqnarray}
c^{(2)}_{ab}(\mathbf{r}_{12},\omega_1,\omega_2,\rho_1,...,\rho_C,T,m_1,...,m_C)=
c_{HS}^{(2)}(r_{12},\rho)+
\nonumber\\
\frac{m_am_b}{\widehat{m}^2}
\left[{c_D^{(2)}(r_{12},{\rho},\widehat{n})D(\omega_{12},\omega_1,\omega_2)+
c_{\Delta}^{(2)}(r_{12},{\rho},\widehat{n})\Delta(\omega_1,\omega_2)}\right],
\label{c2}
\end{eqnarray}
where $\rho=N/V=\sum_{a=1}^C\rho_a$ is the total number density in the volume $V$ of 
the system, $\rho_a=N_a/V$ is the number density of species $a$, $c_{HS}^{(2)}$ 
is the one-component hard sphere direct correlation function, 
while $c_D^{(2)}$ and $c_{\Delta}^{(2)}$ are correlation functions determined by
Wertheim \cite{we} for the one-component dipolar MSA fluid at the same temperature 
but evaluated for an effective dipole moment $m=\widehat{m}$ and at 
an effective packing fraction $\eta=\widehat{\eta}$.  
Accordingly, in Eq. (\ref{c2}) we have introduced
\begin{equation}
\widehat{m}=\sqrt{\frac{\sum_{a=1}^C{m_a^2}}{C}},\,\,\,\,\,\,\,\,\,\,\,\,\,\,\,\,
\widehat{\eta}=\frac{\pi}{6}\sigma^3\frac{\sum_{a=1}^Cm_a^2\rho_a}{\widehat{m}^2}
\label{eta}
\end{equation}
and the rotationally invariant function
\begin{equation}
\Delta(\omega_1,\omega_2)=\widehat{\mathbf{m}}_1(\omega_1)\cdot\widehat{\mathbf{m}}_2(\omega_2).
\end{equation}
In order to explain dependence on $\widehat{n}$ it is convenient to introduce a 
new parameter $\widehat{\xi}\equiv\widehat{\xi}(\widehat{\chi}_{_L})=\widehat{\eta}\,\widehat{n}$ 
which is given by the implicit equation 
\begin{equation}
4\pi\widehat{\chi}_{_L}=q(2\widehat{\xi}\,)-q(-\widehat{\xi}\,),
\label{xi}
\end{equation}
which has the same form as the corresponding equation for the one-component system 
(see Refs. \cite{sd,we}) where
\begin{equation}
\widehat{\chi}_{_L}=\frac{1}{3}\beta\sum_{a=1}^C\rho_am_a^2
\label{ls}
\end{equation}
is the averaged Langevin susceptibility and $\beta=1/({k_BT})$ is the inverse temperature 
with the Boltzmann constant $k_B$. The function $q(x)$ is the reduced inverse 
compressibility function of hard spheres within the Percus-Yevic approximation:
\begin{equation}
q(x)=\frac{(1+2x)^2}{(1-x)^4}.
\end{equation}
For the zero-field (initial) magnetic susceptibility $\chi$ of the mixture the theory by Adelman and Deutch \cite{ad} yields
\begin{equation}
\chi=\frac{\widehat{\chi}_{_L}}{q(-\widehat{\xi}(\widehat{\chi}_{_L})\,)}\,\,\,\,.
\label{sus}
\end{equation}
Concerning Eq. (\ref{c2}) Wertheim \cite{we} and Adelman and Deutch \cite{ad} showed that
\begin{equation}
c^{(2)}_{\Delta}(r_{12},\rho,\widehat{n})
=2\widehat{n}[c_{HS}^{(2)}(r_{12}, 2\widehat{n}\rho)-c_{HS}^{(2)}(r_{12}, -\widehat{n}\rho)],
\label{cm1}
\end{equation}
\begin{equation}
c^{(2)}_{D}(r_{12},{\rho},\widehat{n})=
\overline{c}_D^{(2)}(r_{12},{\rho},\widehat{n})
-3r_{12}^{-3}\int_0^{r_{12}}dss^2\overline{c}_D^{(2)}(s,{\rho},\widehat{n})
\label{cm2}
\end{equation}
with
\begin{equation}
\overline{c}_D^{(2)}(r_{12},{\rho},\widehat{n})=
\widehat{n}[2c_{HS}^{(2)}(r_{12}, 2\widehat{n}\rho)+c_{HS}^{(2)}(r_{12},-\widehat{n}\rho)],
\label{cm3}
\end{equation}
where $c_{HS}^{(2)}(r_{12},\rho)$ is the one-component hard sphere Percus-Yevick 
correlation function at density $\rho$. 
The dimensionless quantity $\widehat{n}=\widehat{\xi}/\widehat{\eta}$ is determined by 
solving Eq. (\ref{xi}).
We find that $\widehat{n}$ vanishes in the nonpolar limit $m_a\rightarrow{0}$ for all $a$. 
This can be inferred from the results of Rushbrooke et al. \cite{rsh} (obtained originally for one-component dipolar fluids) according to which the solution of Eq. (\ref{xi}) can be expressed as a power series in terms of $\widehat{\chi}_{_L}$:
\begin{equation}
\widehat{\xi}=\frac{\pi}{6}\widehat{\chi}_{_L}-\frac{5\pi^2}{48}\widehat{\chi}_{_L}^2+
O(\widehat{\chi}_{_L}^3).
\label{cic}
\end{equation}
From Eqs. (\ref{cic}) and (\ref{eta}) it follows that
\begin{eqnarray}
\lim_{\{m_1,...,m_C\}\rightarrow{0}}\widehat{n}=\nonumber\\
\lim_{\{m_1,...,m_C\}\rightarrow{0}}\frac{\widehat{\xi}}{\widehat{\eta}}=
\lim_{\{m_1,...,m_C\}\rightarrow{0}}
(1+O(\widehat{\chi}_{_L}))\left(\frac{\beta}{3C\sigma^3}\sum_{a=1}^Cm_a^2\right)=0.
\end{eqnarray}
Therefore in this limit the functions $c_D^{(2)}$ and $c_{\Delta}^{(2)}$ vanish 
(see Eqs.$\!$ (\ref{cm1}), (\ref{cm2}), and (\ref{cm3})). Due to $\frac{m_am_b}{\widehat{m}^2}\leq{C}\frac{m_am_b}{m_a^2+m_b^2}\leq{C}$, as expected 
in the nonpolar limit the rhs of Eq. (\ref{c2}) reduces to the direct correlation function $c_{HS}^{(2)}$ of a one-component HS fluid with total density $\rho=\sum_{a=1}^C\rho_a$.

For a one-component system ($C={1}$) the prefactor ${m_am_b/\widehat{m}^2}$ equals 1 
and $\widehat{\eta}=\eta=\frac{\pi}{6}\rho\sigma^3$ so that Eqs. (\ref{xi}) and  (\ref{ls}) 
render $\xi$ for the one-component system which indeed yields 
the one-component direct correlation function. 

Considering the case of a binary mixture of hard spheres and of DHS 
reveals the approximate character of Eq. (\ref{c2}). In this case, for the 
dipolar hard sphere -- hard sphere cross correlations (i.e., $m_a\neq{0}$, $m_b=0$), 
according to Eq. (\ref{c2})  the corresponding correlation function reduces to the a one-component 
hard sphere direct correlation function, which certainly is a rough approximation.
We note that $c^{(2)}_D(r_{12}\rightarrow{\infty})\sim{r_{12}^{-3}}$ is long-ranged (see Eq. (\ref{cm2})) 
while $c^{(2)}_{\Delta}(r_{12}\geq\sigma)=0$ is short-ranged (see Eq. (\ref{cm1})).

In the following we consider DHS mixtures in a homogeneous external magnetic 
field $\mathbf{H}$, the direction of which is taken to coincide with the direction of the $z$ axis. 
For a single dipole the magnetic field gives rise  to the following additional contribution 
to the interaction potential:
\begin{equation}
u^{ext}_i=-\mathbf{m}_i\mathbf{H}=-m_iH\cos\theta_i,
\end{equation}
where the angle $\theta_i$ measures the orientation of the $i$-th dipole relative to the field direction.

\section{Magnetization in an external field}
\label{magn}
In the following we extend our previous theory \cite{sd} to $C$-component and polydisperse 
dipolar mixtures in which 
the particles have the same hard sphere diameter but different strengths of the dipole moments. 
\subsection{Multicomponent systems}
Our analysis is based on the following grand canonical variational functional $\Omega$, which is an extension 
to $C$ components of the one-component functional used in Ref. \cite{sd}:
\begin{eqnarray}
\!\!\!\!\!\!\!\!\!\!\!\!\!\!\!\!\!\!\!\!\!\!\!\!\!\!\!\!\Omega=F_{DHS}[\rho_1,...,\rho_C,
\{\alpha_1(\omega),...,\alpha_C(\omega)\},T]
-\sum_{a=1}^C\rho_a\int{d\omega}\alpha_a(\omega)(\mu_a-u^{ext}_a(\omega)),
\end{eqnarray}
where $F_{DHS}$ is the Helmholtz free energy functional of an anisotropic, dipolar, equally sized 
hard sphere fluid mixture and where $\mu_a$ and $\alpha_a(\omega)$ are the chemical potential 
and the orientational distribution function of the species $a$, respectively. 
Since the external field is spatially constant, $\rho_1,...,\rho_C$ are constant, 
too. Thus $F_{DHS}$ is a function of $\rho_1,...,\rho_C, m_1,...,m_C$ and a functional 
of $\alpha_1(\omega),...,\alpha_C(\omega)$. 
The Helmholtz free energy functional consists of the {\it{ideal}} gas and the {\it{excess}} contribution:
\begin{eqnarray}
F_{DHS}=F^{id}[\rho_1,...,\rho_C,
\{\alpha_1(\omega),...,\alpha_C(\omega)\},T]\nonumber\\
+F^{exc}_{DHS}[\rho_1,...,\rho_C,\{\alpha_1(\omega),...,\alpha_C(\omega)\},T].
\end{eqnarray}
For the $C$-component mixture the ideal gas contribution has the form
\begin{equation}
F^{id}=k_BTV\sum_{a=1}^C\rho_a\left[{\ln(\rho_a\Lambda_a)-1+\int{d\omega}\alpha_a(\omega)
\ln(4\pi\alpha_a(\omega)})\right],
\end{equation}
where $\Lambda_a$ is the de Broglie wavelength of species $a$.
If the system is {\it{anisotropic}} the DHS free energy  $F_{DHS}^{exc,\,ai}$ is approximated by a second-order 
functional Taylor series, expanded around a homogeneous {\it{isotropic}} reference 
system with bulk densities $\rho_1,...,\rho_C$ and an 
isotropic free energy $F_{DHS}^{exc,\,i}$:
\begin{eqnarray}
{\beta}F_{DHS}^{exc,\,ai}[\rho_1,...,\rho_C,\{\alpha_1(\omega),...,\alpha_C(\omega)\},T]=
{\beta}F_{DHS}^{exc,\,i}(\rho_1,...,\rho_C,T)\nonumber\\
-\sum_{a=1}^C\rho_a\int{d^3r_1}{d\omega}\Delta\alpha_a(\omega)
c_{a}^{(1)}(\rho_1,...,\rho_C,T)\nonumber\\
\!\!\!\!\!\!\!\!\!\!\!\!\!\!\!\!\!\!\!\!\!\!\!\!\!\!\!\!\!\!\!
-\frac{1}{2}\sum_{a,b=1}^C\rho_a\rho_b
\int{d^3r_1}{d\omega_1}\int{d^3r_2}{d\omega_2}
\Delta\alpha_a(\omega_1)\Delta\alpha_b(\omega_2)
c^{(2)}_{ab}(\mathbf{r}_{12},\omega_1,\omega_2,\rho_1,...,\rho_C,T),
\label{fex}
\end{eqnarray}
where $\Delta\alpha_a(\omega)=\alpha_a(\omega)-1/(4\pi)$ is the difference between 
the anisotropic ($H\neq{0}$) and the isotropic ($H=0$) orientational distribution function of the component $a$;
$c_a^{(1)}$ and $c_{ab}^{(2)}$ (see Eq.(\ref{c2})) are the first- and second-order direct correlation functions, respectively, of the components of the {\it{isotropic}} DHS mixtures. Since in the isotropic system all $c_a^{(1)}$ are independent of the dipole orientation $\omega$, and because $\int{d\omega}\Delta\alpha_a(\omega)=0$, only the second-order direct correlation functions $c_{ab}^{(2)}$ provide a nonzero contribution to the above free energy functional. Since $c_{ab}^{(2)}$ depends only on the difference vector ${\mathbf{r}}_1-{\mathbf{r}}_2$, Eq.(\ref{fex}) reduces to
\begin{eqnarray}
\!\!\!\!\!\!\!\!\!\!\!\!\!\!\!\!\!\!\!\!
{\beta}F_{DHS}^{exc,\,ai}[\rho_1,...,\rho_C,\{\alpha_1(\omega),...,\alpha_C(\omega)\},T]=
{\beta}F_{DHS}^{exc,\,i}(\rho_1,...,\rho_C,T)\nonumber\\
\!\!\!\!\!\!\!\!\!\!\!\!\!\!\!\!\!\!\!\!\!\!\!\!\!\!\!\!\!\!\!\!
-\frac{1}{2}\rho^2V\sum_{a,b=1}^Cx_ax_b
\int{d\omega_1}\int{d\omega_2}
\Delta\alpha_a(\omega_1)\Delta\alpha_b(\omega_2)
\int{d^3r_{12}\,}c^{(2)}_{ab}(\mathbf{r}_{12},\omega_1,\omega_2,\rho_1,...,\rho_C,T),\nonumber\\
\end{eqnarray}
where $x_a=N_a/N=\rho_a/\rho$ is the mole fraction of the component $a$. 
The expression for the excess free energy ${\beta}F_{DHS}^{exc,i}$ of the isotropic 
DHS system was also given by Adelman and Deutch \cite{ad}.
In the case of a cylindrical sample (elongated around the magnetic field direction ) and 
homogeneous magnetization all $\alpha_a(\omega)$ depend only on the polar angle $\theta$, 
and thus they can be expanded in terms of Legendre polynomials:  
\begin{equation}
\alpha_a(\omega)=\frac{1}{2\pi}\overline{\alpha}_a(\cos\theta)=
\frac{1}{2\pi}\sum_{l=0}^{\infty}\alpha_{al}P_l(\cos\theta),\,\,\,\,a=1,2,...,C\,\,.
\end{equation}
Due to $\alpha_{a0}=1/2$ one has
\begin{equation}
\Delta\alpha_a(\omega)=\frac{1}{2\pi}\sum_{l=1}^{\infty}\alpha_{al}P_l(\cos\theta).
\end{equation}
The second-order MSA direct correlation functions of the DHS fluid mixture 
(see Eq. (\ref{c2})) are used to obtain the excess free energy functional. 
In order to avoid depolarization effects due to domain formation, we consider 
sample shapes of thin cylinders, i.e., needle-shaped volumes $V$. 
Due to the properties of $D$, $\Delta$, and $P_l$ only the 
terms $\alpha_{al}$ with $l\leq1$ contribute to the {\it{excess}} free energy.
Elementary calculation leads to
\begin{equation}
\frac{F_{DHS}^{exc,\,ai}}{V}=f^{exc,\,i}_{DHS}-\frac{2\rho^2}{9{\widehat{\chi}}_{_L}}
(1-q(-\widehat{\xi}\,))\sum_{a,b=1}^Cx_ax_bm_am_b\alpha_{a1}\alpha_{b1},
\end{equation}
where $f_{DHS}^{exc,\,i}=F_{DHS}^{exc,\,i}/V$.
Minimization of the grand canonical functional with respect to the orientational 
distribution functions (note that $f_{DHS}^{exc,\,i}$ does not depend on them) yields
\begin{eqnarray}
\!\!\!\!\!\!\!\!\!\!\!\!\!\!\!\!\!\!
\bar\alpha_a(\omega)=Z_a^{-1}\exp\left(\beta{m_a}
\left(H+\frac{2\rho}{3\widehat{\chi}}_{_L}(1-q(-\widehat{\xi}\,))
\sum_{b=1}^Cx_b{m_b}\alpha_{b1}\right)P_1(\cos\theta)\right),
\end{eqnarray}
with normalization constants $Z_a$ which are fixed by the requirements $\int{d\omega}\alpha_a(\omega)=1$.
With this normalization the expansion coefficients $\alpha_{a1}$ are given by
\begin{equation}
\!\!\!\!\!\!\!\!
\alpha_{a1}=\frac{3}{2}L\left[\beta{m_a}
\left({H+\frac{2(1-q(-\widehat{\xi}\,))}{3\widehat{\chi}_{_L}}\sum_{b=1}^C
\rho_bm_b\alpha_{b1}}\right)\right],\,\,\,\,a=1,2,...,C,
\label{a1}
\end{equation}
where $L(x)=\coth(x)-1/x$ is the Langevin function. Each particle of the magnetic fluid carries a 
dipole moment which will be aligned preferentially in the direction of the external field. 
This gives rise to a magnetization
\begin{equation}
M=\sum_{a=1}^C\rho_am_a\int{d\omega}\alpha_a(\omega)\cos\theta=
\frac{2}{3}\sum_{a=1}^C\rho_am_a\alpha_{a1}.
\label{m1}
\end{equation} 
Equations (\ref{m1}) and (\ref{a1}) lead to an implicit equation for the dependence of the magnetization on the external field:
\begin{eqnarray}
M=\rho\sum_{a=1}^Cm_ax_a
L\left[{\beta}m_a\left(H+\frac{(1-q(-\widehat{\xi}\,))}{\widehat{\chi}_{_L}}M\right)\right].\
\label{mag}
\end{eqnarray}
We note that in Eq. (\ref{mag}) in the limit of weak fields the series expansion of the Langevin 
function, $L(x\rightarrow{0})={x/3}$, reduces to Eq. (\ref{sus}) for the zero-field magnetic susceptibility.
\subsection{Polydisperse systems}
For ferromagnetic grains the dipole moment of a particle is given by
\begin{equation}
m(x)=\frac{\pi}{6}M_s{\mathcal{D}}^3,
\end{equation}
where $M_s$ is the bulk saturation magnetization of the core material and $\mathcal{D}$ is the 
diameter of the particle. Accordingly, our model of equally sized particles with different 
dipole moments applies to systems composed of materials with distinct saturation magnetizations. 
Another possibility consists of considering particles with a magnetic core and a nonmagnetic shell, 
which allows one to vary $m$ via changing the core size with $M_s$ fixed and by keeping the 
overall diameter of the particles fixed via adjusting the thickness of the shell. 
For a small number $C$ of components this can be experimentally realizable.

Equation (\ref{sus}) has been extended even to the description 
of polydisperse ferrofluids \cite{ml,iv}. However, it is unlikely that this extension relates to a realistic 
experimental system because it supposes again that the diameters of all particles are the same.
The two possible realizations mentioned above will be very difficult to implement for a large 
number $C$ of components, mimicking polydispersity. If one nonetheless wants to study such kind 
of a system the expression for its zero-field susceptibility is a natural 
extension of Eq. (\ref{ls}):
\begin{equation}
{\overline{\chi}}_{_L}=\frac{1}{3}\beta\rho\int_0^{\infty}
d\mathcal{D}\,p(\mathcal{D})m^2(\mathcal{D}),
\end{equation}
where $p(\mathcal{D})$ is the probability distribution function for the magnetic core diameter. 
The corresponding zero-field susceptibility of the polydisperse system is
\begin{equation}
{\overline{\chi}}=\frac{{\overline{\chi}}_{_L}}{q(-{\overline{\xi}}(-{\overline{\chi}}_{_L}))},
\label{spoly}
\end{equation}
where $\overline{\xi}$ is the implicit solution of the equation
\begin{equation}
4\pi\overline{\chi}_{_L}=q(2\overline{\xi})-q(-\overline{\xi}).
\end{equation}
We note that similarly the above equation for the magnetization (see Eq. (\ref{mag})) 
can also be extended to polydisperse fluids leading to magnetization curves 
$\overline{M}(H)$ defined implicitly by
\begin{equation}
\overline{M}=\rho\int_{0}^{\infty}\,d{\mathcal{D}}p(\mathcal{D})m(\mathcal{D})
L\left[{\beta}m(\mathcal{D})\left(H+
\frac{(1-q(-\overline{\xi}))}{\overline{\chi}_{_L}}\overline{M}\right)\right].
\label{mag2}
\end{equation}
In the limit of weak fields Eq. (\ref{mag2}) reduces to the expression 
in Eq. (\ref{spoly}) for the zero-field susceptibility. 
In the following we shall do not assess via MC simulations the range of 
validity of Eq. (\ref{mag2}) for polydisperse magnetic fluids, leaving this 
for future studies.
\begin{figure}[t]
\begin{center}
\includegraphics[width=25pc]{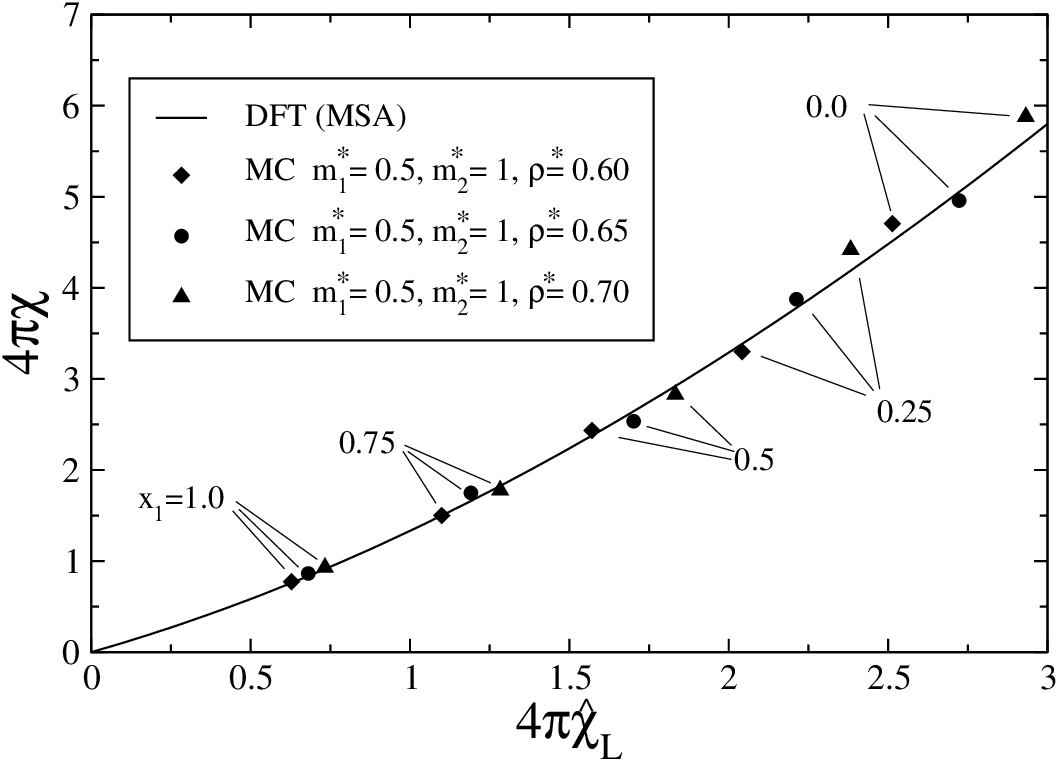}
\caption{Zero-field susceptibility $\chi$ (Eq. (\ref{sus})) as function of the averaged 
Langevin susceptibility $\widehat{\chi}_{_L}\sim{1/T}$ (Eq. (\ref{ls})). 
In terms of these quantities DFT (MSA) predicts the master curve given by the full line. 
The MC data correspond to the following choices of the system parameters. 
The dipole moments are $m_1^*=0.5$ and $m_2^*=1$ in all cases, 
the reduced densities are $\rho^*=0.6$ (diamonds), $\rho^*=0.65$ (circles), 
and $\rho^*=0.7$ (triangles) with the concentrations $x_1=0$, $x_1=0.25$, $x_1=0.5$, 
$x_1=0.75$, and $x_1=1$. The error bars of the MC data are given by the symbol sizes.}
\end{center}
\end{figure} 

\section{Monte Carlo simulations}
In order to assess the predictions of the DFT presented in Sec. \ref{magn} we have carried out 
MC simulations for DHS fluid mixtures using canonical (NVT) ensembles. 
Boltzmann sampling and periodic boundary conditions with the minimum-image 
convention \cite{al} have been applied. 
A spherical cutoff of the dipole-dipole interaction potential at half of the linear 
extension of the simulation cell has been applied and the reaction field long-ranged 
correction \cite{al} with a conducting boundary condition has been adopted. 
For obtaining the magnetization data, after 40.000 equilibration cycles 0.8-1.0 million 
production cycles have been used involving 1024 particles. In the simulations with an 
applied field the equilibrium magnetization is obtained from the equation
\begin{equation}
\mathbf{M}=\frac{1}{V}\left\langle{\sum_{i=1}^N{\mathbf{m}}_i}\right\rangle,
\end{equation}
\begin{figure}[t]
\begin{center}
\includegraphics[width=25pc]{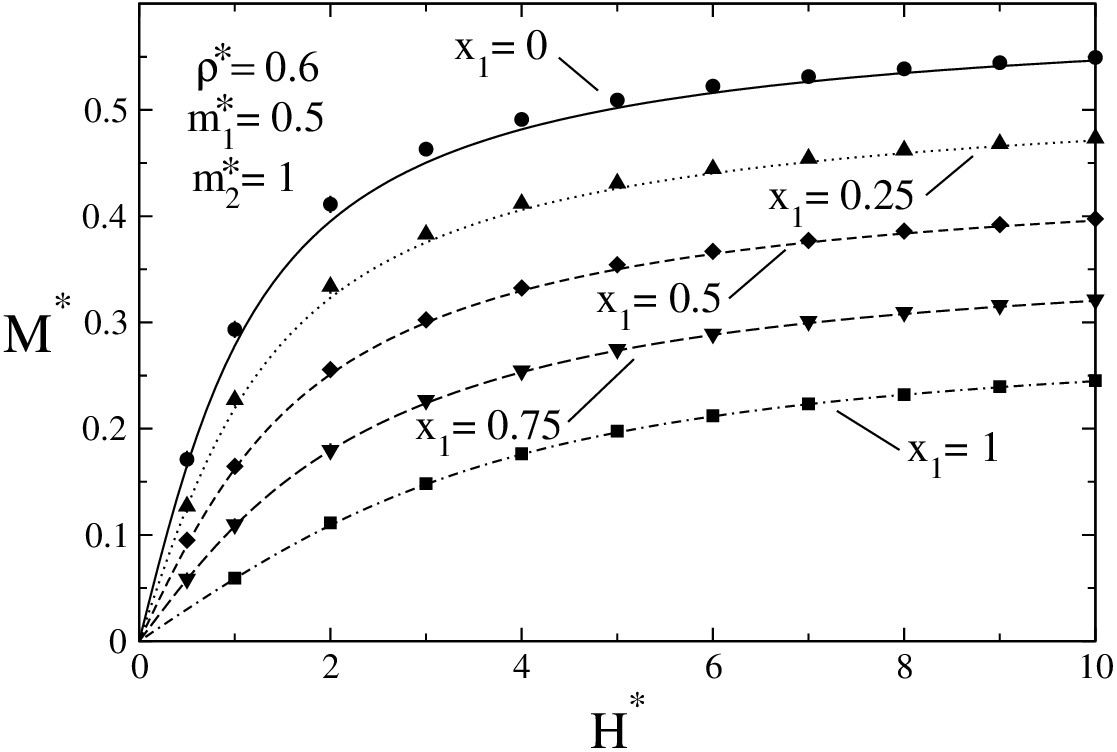}
\caption{Magnetization curves of a binary DHS fluid mixture for five values of the 
concentration $x_1$ of the species with $m^*_1=0.5$; $m_2^*=1$. 
The overall number density is $\rho^*=0.6$. 
The curves correspond to the predictions of the MSA based DFT (Eq. (\ref{mag})). 
The symbols are MC simulation data. Their error bars are given by the symbol sizes.}
\end{center}
\end{figure} 
where the brackets denote the ensemble average. In simulations without external field the zero-field 
magnetic susceptibility has been obtained from the corresponding fluctuation formula
\begin{equation}
\chi=\frac{\beta(\langle{{\mathbf{\mathcal{M}}}^2}\rangle-\langle{{\mathbf{\mathcal{M}}}}\rangle^2)}{3V},
\end{equation}
where $\mathcal{M}=\sum_{i=1}^N{\mathbf{m}}_i$ is the instantaneous magnetic dipole 
moment of the system. Statistical errors have been determined from the standard deviations of 
subaverages encompassing 100.000 MC cycles.

\section{Numerical results and discussion}

In the following we shall use reduced quantities: $\rho^*=\rho\sigma^3$ as the reduced density, 
$m^*_a=m_a/\sqrt{k_BT\sigma^3}$ as the dimensionless dipole moment of species $a$, 
$H^*=H\sqrt{\sigma^3/(k_BT)}$ as the reduced magnetic field strength, 
and $M^*=M\sqrt{\sigma^3/(k_BT)}$ as the reduced magnetization. 
The  calculation of the zero-field susceptibility and the magnetization $M(H)$ 
of the multicomponent DHS fluid mixtures (with identical particle diameters but 
different dipole moments) can be summarized by the sequence of the following steps: \newline
1) calculation  of the Langevin susceptibility according to Eq. (\ref{ls}),\newline
2) solving Eq. (\ref{xi}) for $\widehat{\xi}$,\newline
3) calculation of the zero-field susceptibility according to Eq. (\ref{sus}),\newline
4) calculation of the magnetization for a given value of $H$ according to Eq. 
(\ref{mag}) using the consecutive approximation method with $M=0$ as initial value. 
The convergence of this consecutive approximation is very good, obtaining the limiting 
results within 5-8 cycles.\newline
\begin{figure}[t]
\begin{center}
\includegraphics[width=25pc]{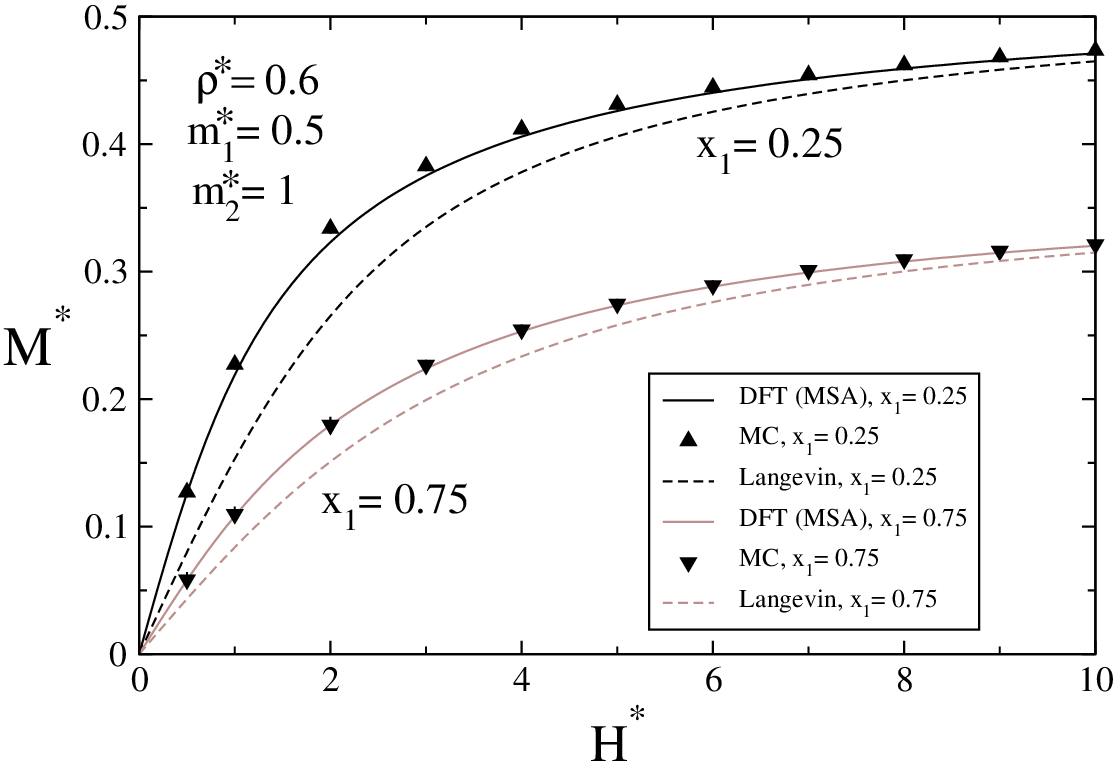}
\caption{Same system as in Fig. 2. For $x_1=0.25$ and $x_1=0.75$ DFT (MSA) and the 
corresponding MC data are compared with the Langevin magnetization curves. 
The error bars of the MC data are given by the symbol sizes.}
\end{center}
\end{figure} 
Figure 1 shows the dependence of the zero-field susceptibility $\chi$ on the 
Langevin susceptibility $\chi_{_L}\sim{1/T}$ (see Eq. (\ref{ls})) for dipolar 
hard sphere mixtures as obtained from Eq. (\ref{sus}) and from the numerical solution of 
Eq. (\ref{xi}). This result is compared with MC simulation data for a 
binary mixture $[(m_1^*,m_2^*)=(0.5,1)]$ for three total number 
densities $\rho^*$ and five concentrations $x_1$. 
In these cases DFT (MSA) provides a good approximation for the initial susceptibility of 
this binary system within the range $0\leq{4\pi}\chi_{_L}\lesssim2.5$. 
Within this range the two-component system with various concentrations and densities can be 
described by the same master curve which is the same also for 
different systems $(m_1^*,m_2^*,...,m_C^*)$.  This is the main statement of the 
MSA theory by Adelman and Deutch \cite{ad}.

Figure 2 displays the magnetization curves of the two-component DHS fluid 
mixture with $(m_1^*,m_2^*,\rho^*)=(0.5,1,0.6)$ for five values of the concentration. 
For high values of $H^*$ we find excellent, quantitative agreement for 
all concentrations between the DFT (MSA) results and the MC data. 
For small values of $H^*$, especially for $H^*=0.5$, i.e., in the linear response regime, 
the agreement between the simulation data and the DFT results is also very good, 
 which matches with the good agreement found for the zero-field susceptibility (see Fig. 1). 
Close to the elbow of the magnetization curves the level of quantitative agreement 
reduces significantly for smaller concentrations $x_1$ of the magnetically weaker component, 
while it remains good for large concentrations $x_1$. 
We note that also for two-dimensional systems this range is the most 
sensitive one concerning the agreement between theoretical results 
and simulation data \cite{ks}. For the same system  Fig. 3 displays 
a comparison between the DFT results together with the MC data 
and the corresponding Langevin theory. 
This shows that the interparticle interaction enhances the magnetization 
relative to the corresponding values of the Langevin theory. 
For a three-component DHS fluid mixture [$(m_1^*,m_2^*,m_3^*)=(0.5,0.75,1)$] Fig. 4 displays the dependence of the magnetization on the concentration $x_3$ for a fixed field strength $H^*=2$ and for three values of the concentration $x_1$; $x_1+x_2+x_3=1$. 
Since the value $H^*=2$ falls into the aforementioned elbow 
regime of the magnetization curves, for the small concentration $x_1=0.25$ of the less polar ($m_1^*=0.5$) component 
the DFT (MSA) results underestimate the simulation data. 
At higher concentrations of $x_1$ the agreement between DFT and the simulation 
data is much better. This is expected to occur, because for large $x_1$ the fluid is 
dominated by less polar particles.
\begin{figure}[t]
\begin{center}
\includegraphics[width=25pc]{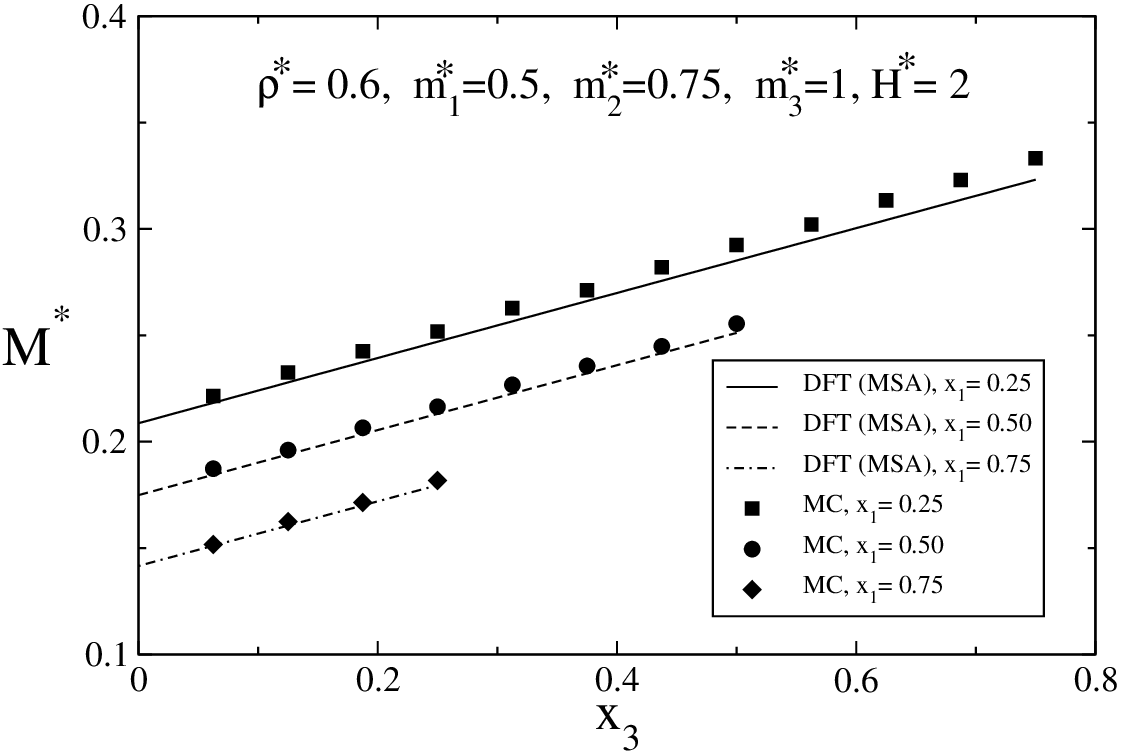}
\caption{Dependence of the magnetization on the concentration $x_3$ of 
the species with $m_3^*=1$ for a three-component 
DHS fluid mixture for a fixed field strength $H^*=2$ and concentrations 
$x_1=0.25$, $x_1=0.5$, and $x_1=0.75$; $x_1+x_2+x_3=1$. 
The reduced dipole moments of the species 1 and 2 are 
$m_1^*=0.5$ and $m_2^*=0.75$. The overall number density is $\rho^*=0.6$. 
The error bars of the MC data are given by the symbol sizes.}
\end{center}
\end{figure} 
\section{Summary}
We have obtained the following main results:\\
1) Based on a second-order Taylor series expansion of the anisotropic free energy 
functional of equally sized dipolar hard spheres with different dipole moments 
and by using the mean spherical approximation (MSA) we have derived an 
analytical expression (Eq. (\ref{mag})) for 
the magnetization of multicomponent ferrofluidic mixtures in external fields. 
This implicit equation extends the applicability of MSA to the presence of 
external magnetic fields of arbitrary strengths. 
We find quantitative agreement between the 
results from this DFT (MSA) and our Monte Carlo simulation data for Langevin 
susceptibilities $4\pi\chi_{_L}\lesssim{2.5}$ (Figs. 2, 3, and 4).\\
2) As confirmed also by MC simulation data the zero-field susceptibility of multicomponent ferrofluids can be expressed by 
a single master curve in terms of the Langevin susceptibility (Fig. 1). 
Beyond the linear response regime the magnetization curves of multicomponent ferrofluids 
cannot be reduced to a single master curve (Eq. (\ref{mag})).\\  
3) By applying the MSA theory for the magnetic susceptibility to polydisperse 
systems we have extended the multicomponent magnetization equation to polydisperse systems.
  
\section*{Acknowledgments}
I. Szalai acknowledges financial support for this work by the Hungarian State and the European 
Union within the project TAMOP-4.2.1/B-09/1/KONV-2010-0003.

\end{document}